\documentclass[10pt,letterpaper,twocolumn]{article} 

\usepackage{ol2}
\usepackage[draft]{hyperref}
\usepackage{amsmath}
\usepackage{amssymb}

\begin{document}

\twocolumn[ 

\title{Measurement of scaling laws for shock waves in thermal nonlocal media}


\author{ N. Ghofraniha$^{1*}$, L. Santamaria Amato $^2$, V. Folli$^2$, S. Trillo$^3$, E. DelRe$^{4,1}$, and C. Conti$^{4,2}$}
\address{
$^1$IPCF-CNR, UOS Roma Kerberos, Universit\`{a} La Sapienza, P. le A. Moro 2,
I-00185, Roma, Italy\\
$^2$ISC-CNR, UOS Sapienza, P. A. Moro 
2, 00185 - Roma, Italy\\
$^3$Dipartimento di Ingegneria, Universit\`{a} di Ferrara, Via Saragat 1, 44100 Ferrara, Italy\\
$^4$Dipartimento di Fisica - Universit\`{a} La Sapienza, P. A. Moro 
2, 00185 - Roma, Italy\\
$^*$Corresponding author: neda.ghofraniha@roma1.infn.it
}

\begin{abstract}
We are able to detect the details of spatial optical collisionless wave-breaking
through the high aperture imaging of a beam suffering shock in a fluorescent nonlinear nonlocal thermal medium.
This allows us to directly measure how nonlocality and nonlinearity affect the point of shock formation
and compare results with numerical simulations.
\end{abstract}


 ] 

\noindent
Shock waves (SW) are characterized by an abrupt wave-profile, corresponding to discontinuous solutions of hyperbolic
partial differential equations~\cite{whitman}.
{\it Dissipative} SW display a smooth wave front,  due to viscous damping, whereas
{\it dispersive}, or collisionless, SW exhibit a characteristic oscillatory front. 
The latter are expected in Hamiltonian universal models
for nonlinear media, such as the nonlinear Schr\"{o}dinger (NLS) equation,
in the weakly dispersive regime where hydrodynamical approximations hold true~\cite{gurevich, bronski, kamchatnov}. 
Dispersion regularizes the shock, determining the onset of oscillations that appear near the wavebreaking
point, and expand from there \cite{gurevich}, as investigated in ion-acoustic waves ~\cite{taylor},  in fibers \cite{rothenberg,conti10},  in diffracting optical beams \cite{wan} or in  Bose-Einstein condensates \cite{hoefer}.
Nonlocal phenomena \cite{deykoon, wyller, conti, wurtz,con10} 
substantially alter dispersive SW~\cite{barsi, ghofraniha,conti09}.
For propagation  in highly nonlocal, strongly defocusing, thermal media,
previous work has been based on approximating the 3D dimensional Fourier equation for the temperature
profile by a nonlocal nonlinearity with a finite degree of nonlocality \cite{ghofraniha,Minovich2007,conti09}.

In this Letter we report on the quantitative investigation of the formation of dispersive shocks from standard  
laser beams (Gaussian TEM$_{00}$) in thermal media.
The key to our experiments is the direct observation of the intensity profile of the shock wave in its formation 
and evolution through the use of a fluorescent medium and a high aperture microscope. We quantify the shock 
position along propagation and show how it depends on nonlinearity and nonlocality.
Our results not only confirm the finite degree of nonlocality of thermal media but actually provide a strategy for its direct measurement.

In experiments we use aqueous solutions of RhodamineB (RhB) displaying a thermal defocusing effect due to light absorption.
A  continuous-wave solid state laser  at wavelength $\lambda$=532~nm is focused  on the input facet of the sample 
at the beam radius waist  $w_0$. 
We work at  two different $w_0$ (18$\mu$m and and 25$\mu$m) to change both nonlinearity and  nonlocality and two different RhB concentrations  
(0.1 mM, sample A; 0.067 mM, sample B) that vary the nonlinearity.
For the visualization of light along its propagation, samples are put in 1cm$\times$1cm$\times$3cm glass cells
(propagation along $1$cm), and top images  are collected  through a microscope and recorded by a 1024$\times$1392 pixel CCD camera.
Conversely, to detect light transmitted at the exit of the samples,  we use 1mm$\times$1cm$\times$3cm glass cells
with propagation along the $1$mm  vertical direction (parallel to  gravity) so as to moderate the effect of heat convection.
Transverse  images of the beam intensity distributions at the exit facet of the cell  are collected through a lens 
and  recorded by the CCD camera.  Images from 
the two experimental configurations are reproduced with comparable length-scale.
 
\begin{figure}[h]
\centerline{\includegraphics[width=8.3cm]{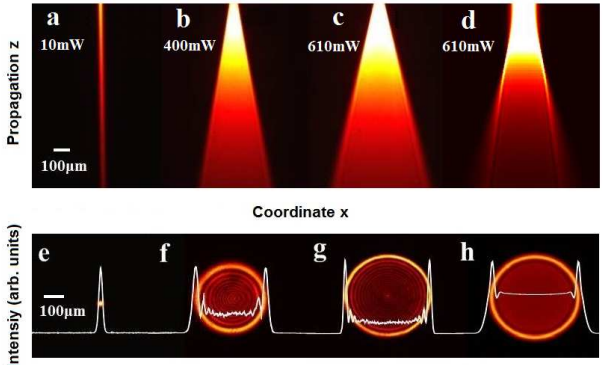}}
\caption{\label{fig1} (Color online)  (a-d) Beam propagation as observed from top fluorescence emission for
three different input powers (a-c) and calculated by numerical simulation (d);
(e-g) corresponding transmitted intensity in the experiments and (h) in the simulations. 
Superimposed curves give the intensity x-profile.
} 
\label{imagshock}
\end{figure}

\begin{figure}[ht]
\includegraphics[width=8.3cm]{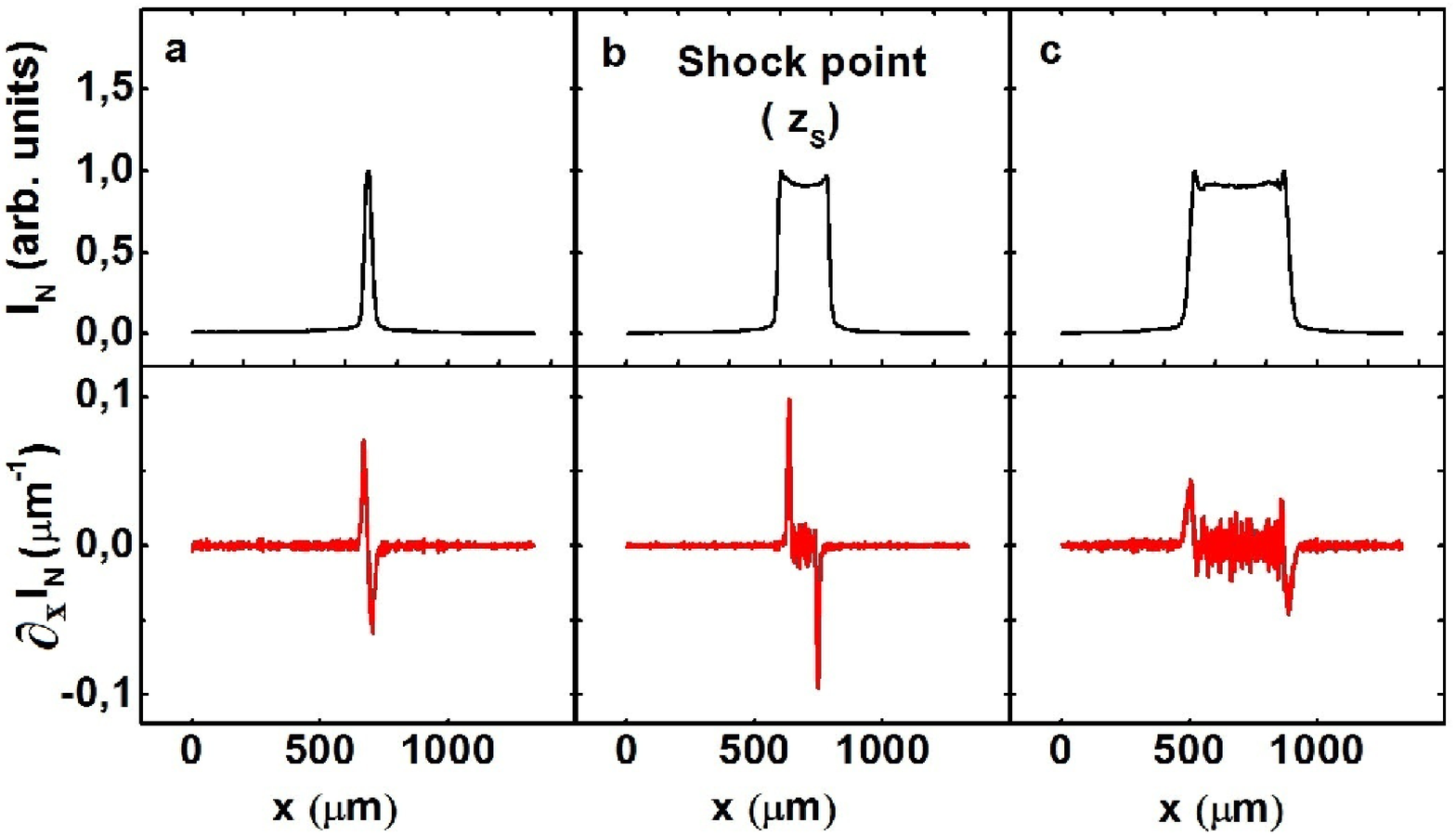}
 \caption{\label{fig2}(Color online)  Normalized intensity profiles  $I_N$ (black line) along with derivative ($\partial_x I_N$, red line) for
three different $z$ at P=600mW: (a) input, $z=0$; 
(b) $z=z_s=273 \mu$m, shock point; (c) $z=850 \mu$m, post-shock.}
\label{derivata}
\end{figure}

\begin{figure}[ht]
\includegraphics[width=7.8cm]{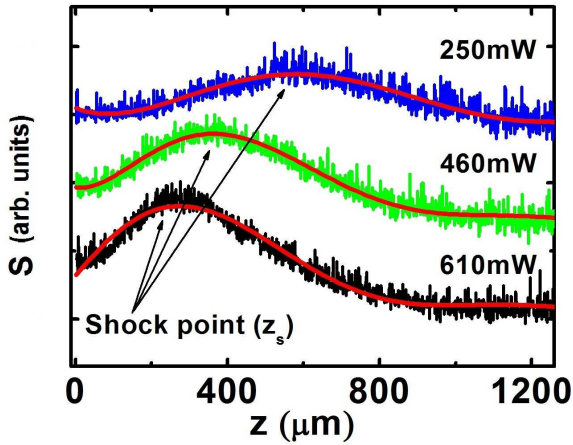}
\caption{\label{fig3} (Color online) Steepness $S(z)$ of transverse profiles as function of $z$ for three different powers. 
The point of maximum corresponds to shock distance $z_s$.
} 
\label{maxder}
\end{figure}

\begin{figure}[h!]
\includegraphics[width=8.3cm]{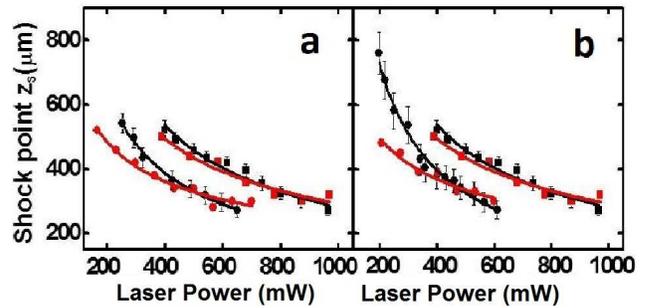}
 \caption{\label{fig4}(Color online)   (a) Shock position $z_s$ versus input power for sample B at two different beam  waists:  
$\blacksquare$ 25$\mu$m , {\Large$\bullet$} 18$\mu$m;
(b)  Shock position vs.  input power for beam waist=25$\mu$m and different dye concentrations: 
$\blacksquare$ 0.067 mM, { \Large$\bullet$} 0.1 mM. 
Experiments are denoted by black (dark)  and simulations by red (clear) symbols.} 
\label{shockpoint}
\end{figure}

Numerical simulations are also performed to verify the experimental measurements and to clarify the role of 
nonlocality. By using a split-step beam propagation method, we numerically integrate the following
nonlocal NLS equation (details in \cite{ghofraniha,conti09}):
\begin{equation}
 \label{NLS1}
i\frac{\partial\psi}{\partial \zeta}+\frac{1}{2}\nabla_\bot^2\psi-\theta\psi=-i\frac{\alpha}{2}\psi,\hspace{0.1cm}
-\sigma^2\nabla_\bot^2\theta+\theta=|\psi|^2,
\end{equation}
where $\psi=\sqrt{\chi L_d} E$ is the normalized field envelope ($I\equiv|E|^2$ is the intensity), 
$\theta=k_0\Delta n L_{d}$, with $\Delta n$ the nonlinear refractive index, $L_d=k_0 n_0 w_0^2$ the
diffraction length, $\chi=k_0 n_2$, $k_0=2\pi/\lambda$, and $n_0$ the refractive index. 
$\alpha=\alpha_0 L_d$ is the normalized loss coefficient defined in terms of the power absorption coefficient $\alpha_0$. 
Transverse spatial coordinates are scaled as  $\xi,\eta=x/w_0,y/w_0$, while $\zeta=z/L_d$. 
Here $\sigma$ is the \textit{degree of nonlocality}, used as a fitting parameter below.
Figures~\ref{imagshock}a-c show three collected images of the transverse distribution of the beam intensity along $z$,
for $w_0=25\mu$m and three beam powers $P$ in sample A.
In Fig.~\ref{imagshock}a, corresponding to the lowest $P$, the beam shape does 
not change (diffraction is weak) along propagation and  the shock does not occur.
The presence of oscillatory SW can be observed in the other two images reported in
 Fig. \ref{imagshock}b and Fig. \ref{imagshock}c at higher powers,
and in Figs.~\ref{imagshock}e-g where the 2-D profiles of the transmitted 
beam  (on $x$-$y$ plane) at the exit facet are reported. The $x$-profiles display post-shock rings
that increase in number and visibility with $P$, with outer rings being more intense than the inner ones,
as typical for dispersive SW from Gaussian beams in nonlocal media.  
Figures~\ref{imagshock}d and \ref{imagshock}h present beam propagation and transmission results obtained from simulations,
performed  by using input parameters  identical to those in experiments and $\sigma=0.2$.
Shock occurs in the regime $L_{nl} \ll L_{\alpha}, L_{d}$ \cite{gurevich, bronski, kamchatnov}, 
with $L_{\alpha}=1/\alpha_0$  the absorption length, and $L_{nl}=1/\chi I(x,0)_{peak}$ is the nonlinear length
(at the highest powers in our experiment, $L_{nl}/L_{\alpha} \sim 0.002$ and, in terms of the smallness 
parameter defined in Refs. \cite{ghofraniha,conti09}, $\varepsilon=\sqrt{L_{nl}/L_d} \sim 0.015$). 
This means that the involved length scales are such that the beam is mainly affected by the nonlinear defocusing 
effect rather than by absorption and diffraction.
In this regime, as a first approximation, the beam phase depends on the intensity profile as $\phi(\xi,\eta,\zeta)=\frac{k_0\zeta}{n_0}\Delta n(I(\xi,\eta))$. 
As discussed in Ref. \cite{ghofraniha}, SW originates in correspondence with the chirp singularity $|d\phi/d\xi|\rightarrow\infty$ 
from which the shock distance can be analytically determined in the hydrodynamical approximation. 
For a generic degree of nonlocality and by assuming, for the sake of simplicity, a one-dimensional generic response 
function $K(\xi-\xi')$ yielding the index change as $\Delta n=\int d\xi' K(\xi-\xi')I(\xi',\eta)$ 
[Eqs. (1) imply a 2-D Lorentzian kernel $K$ as one can easily verify by taking the Fourier transform of the second equation], 
we have:
\begin{equation}
 \label{derphi}
\frac{\partial\phi}{\partial \xi}=\frac{k_0\zeta}{n_0}\int d\xi' K(\xi-\xi')\frac{\partial I}{\partial \xi'},
\end{equation}
and the maximum of $\operatorname{grad}(\phi)$ along $z$ (or $\zeta$) gives the shock position.
Since $K$ is $z$-independent the shock-point $z_s=\zeta L_d$ 
can be directly estimated both in experiments and simulations as the point of largest steepness of  beam intensity, as detailed below.
We note that in the highly nonlocal regime, $K(\xi)=K_0$, and correspondingly $\partial_\xi\phi=(k_0\zeta K_0 / n_0) \int_\xi \partial_\xi I=0$,
i.e., as the degree of nonlocality increases, the response function filters out all the edge frequencies in the broadened 
intensity spectrum in proximity of the shock-point, thus delaying it. In the highly nonlocal limit, the mean-field 
approximation fully smooths out all the singularities and the shock disappears. 
 To estimate the point of shock formation both in experiments and simulations, we calculate the 
$x-$derivative $\partial_x I_{N}$ of the intensity normalized to its peak value, say $I_N(x,z)$. 
We define the steepness $S(z)$ as the maximum over $x$ (at any $z$) of the derivative  $S(z)=\max_x[\partial_x I_{N}(x,z)]$.
Then the shock distance $z_s$ is defined as the distance $z$ corresponding to the maximum steepness,  
[$\max_z(S(z))$  $\rightarrow z=z_s$].  
We follow this procedure for the whole set of images. In Fig.~\ref{derivata} we show representative profiles along with 
their derivatives at the input (Fig.~\ref{derivata}a), right at the shock position (Fig.~\ref{derivata}b), and beyond $z_s$ (Fig.~\ref{derivata}c), the steepness resulting 
the highest in Fig.~\ref{derivata}b where the shock occurs. 
The behavior of the steepness $S$ along $z$ as obtained from experimental data is illustrated in Fig.~\ref{maxder} for three power values.
The shock distance  $z_{\textrm{s}}$ is given by the position of the peak of the ten-degree polynomial fitting curve
(solid curves through data).  In Fig.~\ref{maxder} it is clear that this distance is shorter for increasing power, 
meaning that larger laser intensities speed up the shock formation process.
We report in Fig. \ref{shockpoint}  the measured shock distance versus laser power, comparing with simulations.
In particular Fig.  \ref{shockpoint}a compares the shock point behavior for two beam waists in the same sample (B), 
whereas Fig. \ref{shockpoint}b compares the behavior in the two samples at a given beam waist (25$\mu$m).  
These data show quantitatively how both increasing input intensity or dye concentration speeds up shock formation because both result in a larger nonlinear response.
In the simulations, we make use of the measured values of the physical parameters: 
$n_0=1.3$, $\alpha_0$= 900 (A) and 600 (B) m$^{-1}$;   $\alpha_{\mathrm NL}$= 1.3$\times10^{-5}$  (A) and  1.2$\times10^{-6}$ (B)   
m W$^{-1}$ ;   $n_2=  4\times10^{-11}$  (A) and $2.7\times10^{-11}$ (B)   m$^2$ W$^{-1} $, while $\sigma$ is used as a free fitting parameter.
The best-fit of numerical simulations with the experimental data in Fig.  \ref{shockpoint}a, yields a degree of nonlocality $\sigma=0.63$ ($w_0=18 \mu$m) and $0.3$ ($w_0=25 \mu$m). Following Ref. \cite{ghofraniha}, the nonlocality degree should scale with the inverse of beam waist. The observed slight discrepancy is attributed to the theoretical two-dimensional heat equation used to model the actual temperature profile in our sample.
It is known from the hydrodynamic limit that the shock distance scales with power $P$ according to the law $z_s\propto P^{-\gamma}$, with $\gamma=0.5$ \cite{ghofraniha,conti09}. This power law is confirmed by our experiments and simulations of Eqs. (\ref{NLS1}), though with  deviations in the exponent.
The experimental and numerical data given in Fig.~\ref{shockpoint}, once fitted with $\gamma$ as a free parameter,
yield the following values: on sample A,  $\gamma=0.88$  ($w_0=25\mu$m), $\gamma=1.0$   ($w_0=18\mu$m)  
and on sample B, $\gamma=0.70$  ($w_0=25\mu$m),   $\gamma=0.72$   ($w_0=18\mu$m). 
The values of $\gamma$ from the simulations are: on sample A,  $\gamma=0.43$ ($w_0=25\mu$m), on sample B:   
$\gamma=0.57$ ($w_0=25\mu$m) and  $\gamma=0.43$ ($w_0=18\mu$m).
The discrepancy between simulations and experiments is, once again, ascribed to the limits of validity of the 2D reduction of the heat equation adopted in Eqs. (\ref{NLS1}).

In conclusion, we use fluorescence imaging to directly monitor spatial wave-breaking.
The full intensity distribution map allows us to experimentally determine the shock position for different 
nonlinearity and nonlocality. The comparison with numerical simulations of the nonlocal NLS model enables to quantify the limits of validity of the 2D reduction of the heat equation. 

The research leading to these results has received funding from the European Research Council under the European Community's Seventh
Framework Program (FP7/2007-2013)/ERC grant agreement n. 201766,  from the Italian Ministry of Research (MIUR) through the PRIN project no.2009P3K72Z and from  the Italian Ministry of Education, University
and Research under the Basic Research Investigation
Fund (FIRB/2008) program/CINECA grant code
RBFR08M3P4 and RBFR08E7VA. We thank M. Deen Islam for the technical assistance.



\end{document}